\documentstyle[aps,prl,preprint,floats,epsfig]{revtex}

\begin{document}

\preprint{\tighten\vbox{\hbox{\hfil CLNS 02/1793}
                        \hbox{\hfil CLEO 02-11}
}}

\title{Measurement of Exclusive $B$ Decays to Final States Containing a Charmed Baryon}
\author{CLEO Collaboration}
\date{July 31, 2002}

\maketitle
\tighten

\begin{abstract} 
Using data collected by the CLEO detector in the $\Upsilon(4S)$ region, 
we report new measurements of the exclusive decays of $B$ mesons into final   
states of the type $\Lambda_c^+\overline{p}n(\pi)$, where
$n=0,1,2,3$. We find signals in modes with one, two and three pions and an upper limit for the two body 
decay $\Lambda_c^+\overline{p}$. We also make the first measurements of 
exclusive decays of $B$ mesons to $\Sigma_c\overline{p}n(\pi)$, where
$n=0,1,2$. We find signals in modes with one and two pions and an
upper limit for the two body decay $\Sigma_c\overline{p}$. 
Measurements of these
modes shed light on the mechanisms involved in $B$ decays to baryons.

\end{abstract}
\newpage

{
\renewcommand{\thefootnote}{\fnsymbol{footnote}}


\begin{center}
S.A.~Dytman,$^{1}$ J.A.~Mueller,$^{1}$ S.~Nam,$^{1}$
V.~Savinov,$^{1}$
S.~Chen,$^{2}$ J.~W.~Hinson,$^{2}$ J.~Lee,$^{2}$
D.~H.~Miller,$^{2}$ V.~Pavlunin,$^{2}$ E.~I.~Shibata,$^{2}$
% ...
\end{center}
 
\small
\begin{center}
$^{1}${University of Pittsburgh, Pittsburgh, Pennsylvania 15260}\\
$^{2}${Purdue University, West Lafayette, Indiana 47907}\\
% ...
\end{center}

\setcounter{footnote}{0}
}
\newpage


A distinctive feature of the $B$ system is that the large mass of the $b$-quark allows 
weak decays of the $B$ mesons to proceed via the creation of a baryon anti-baryon pair. 
The mechanisms for baryon production have been the subject of several studies in the last
decade. The dominant decay mechanism is expected to be 
$b \rightarrow c\overline{u}d$ transitions
via internal or external $W$-decay. This can lead to final states including a 
charm meson as well
as a baryon-antibaryon pair, as recently observed by CLEO\cite{TONY} and Belle\cite{BELLE}. 
However, the simplest
decay diagrams lead to states of the form
$\overline{B}\rightarrow\Lambda_c\overline{N}X$, where $\overline{N}$
represents an anti-nucleon. Charge-conjugate processes are implied 
throughout this paper. Inclusive
studies of $\Lambda_c^+$ production from $B$ decays 
indicate a branching fraction of around
5\%, and the soft $\Lambda_c^+$ momentum spectrum indicates that multi-body decays 
dominate\cite{ARGUS,CLEO}. 
In 1996, CLEO made the first exclusive measurements\cite{CLEOO} 
of decays of this type, and found
${\cal B}({B}^-\rightarrow\Lambda_c^+\overline{p}\pi^-)=
(0.62^{+0.23}_{-0.21}\pm0.11\pm0.10)\times 10^{-3}$
and
${\cal B}(\overline{B}^0\rightarrow\Lambda_c^+\overline{p}\pi^+\pi^-)=
(1.33^{+0.46}_{-0.42}\pm0.31\pm0.21)\times 10^{-3}$.
The analysis presented here uses a larger CLEO data sample and improved analysis
techniques to make further measurements of this type, confirming the 
previous observations and measuring
new modes ${B}^-\rightarrow\Lambda_c^+\overline{p}\pi^+\pi^-\pi^-$
and 
${B}^-\rightarrow\Lambda_c^+\overline{p}\pi^-\pi^0$.
Furthermore, by investigation of the resonant substructure of these decays, 
the first exclusive decays of the form
$\overline{B}\rightarrow\Sigma_c\overline{N}X$ have been measured. 
Comparisons of the branching fractions of these 
modes gives information on the underlying mechanisms involved.

The data were collected with two detector configurations, CLEO II\cite{KUB} and 
CLEO II.V\cite{HILL}, 
at the Cornell Electron Storage Ring. 
The data comprise 9.17 ${\rm fb}^{-1}$ taken at the
$\Upsilon(4S)$ which corresponds to $9.74\times 10^6 B\overline{B}$ pairs,
together with 4.6 ${\rm fb}^{-1}$  
taken in the $e^+e^-$ continuum below the $\Upsilon(4S)$ that are used to 
evaluate possible backgrounds.
We assume that the
produced $B^+B^-$ rate is the same as $B^0\overline{B^0}$ at the $\Upsilon(4S)$.

The signal $B$ meson candidates are fully reconstructed by combining detected photons, 
protons, and charged kaons and pions. The tracking system consisted of several 
concentric detectors operating
inside a 1.5 T solenoid. For CLEO II, 
the tracking system consisted of a 6-layer straw tube
chamber, a 10-layer precision drift chamber, and a 
51-layer main drift chamber. The main 
drift chamber also provided a measurement of the specific
ionization, $dE/dx$, used for particle identification. 
For CLEO II.V, the straw tube chamber was replaced
by a 3-layer, double-sided silicon vertex detector, 
and the gas in the main drift chamber was changed from 
argon-ethane to helium-propane mixture.
In both configurations, photons were detected by an 
electromagnetic calorimeter consisting of 
7800 Thallium-doped CsI crystals.

$\Lambda_c^+$ candidates are reconstructed in the modes
$pK^-\pi^+, pK^0_S$ and $\Lambda\pi^+$, where $K^0_S\rightarrow\pi^+\pi^-$ and 
$\Lambda\rightarrow p\pi^-$. 
The $K^0_S$ and $\Lambda$ candidates
are reconstructed from oppositely charged tracks which form a vertex
well detached from the main event vertex in the plane transverse to the beam direction.  
The invariant mass of the $K^0_S(\Lambda)$ candidate 
is required to be within 8 (3.5) ${\rm MeV}/c^2$ of the known mass. Neutral pion candidates
are formed from pairs of showers detected in the calorimeter 
which yield a $\gamma\gamma$ invariant 
mass within 2 standard deviations of the known $\pi^0$ mass.
The $\Lambda$, $K^0_S$ and $\pi^0$ candidates were then all kinematically 
constrained to their known masses.

Particle identification of $p,K^-$, and $\pi^+$ candidates was performed
using specific ionization measurements in the drift chamber,
and when present, time-of-flight measurements.
For each mass hypothesis, a combined $\chi^2$ probability $P_i$ was formed
($i=\pi,K,p$). Using these $P_i$'s, a normalized
probability ratio $L_i$ was evaluated, where
$L_i=P_i/(P_{\pi}+P_K+P_p)$. Real protons have
$L_p$ of close to 1.0, 
whereas tracks due to other particles are concentrated 
near $L_p=0.0$. For a track to be used as a proton daughter of a 
$\Lambda_c^+\rightarrow pK^-\pi^+$
or $pK^0_S$ 
decay, we require it to have
$L_p>0.9$, which eliminates much of the background but with considerable diminution of
efficiency. For kaons we applied a slightly looser and more efficient
requirement of $L_K>0.7$. We have chosen these selection criteria 
using a Monte Carlo simulation
program to maximize the significance of the $\Lambda_c^+$ signals.
The candidate anti-proton which is the direct daughter of the $B$ meson has a looser 
requirement
of $L_p>0.1$ as it has a higher momentum distribution and lower backgrounds than the 
decay daughter of the $\Lambda_c^+$. The proton from the 
$\Lambda\rightarrow p\pi^-$, and all the 
pion candidates, are required to have particle identification parameters consistent with
their hypothesis. Tracks with no particle identification information are assumed to be
due to pions.

To suppress continuum background, the normalized Fox-Wolfram second moment \cite{WOKF}
is required to be less than 0.35. The number of $\Lambda_c^+$ candidates from the 
$\Upsilon(4S)$ data, after the contribution from continuum events is accounted for
by a subtraction of scaled continuum data, is $7100\pm350\ (12100\pm450)$ 
from the CLEO II (CLEO II.V)
detector configurations.

To reconstruct exclusive $B$ decays we select $\Lambda_c^+$ candidates whose mass
is within 2$\sigma$ of the nominal mass. The mass resolution, $\sigma$,
was calculated
for each of the three decay modes and two detector configurations separately by use of a 
GEANT-based Monte Carlo simulation program\cite{GEANT}. 
We constrain the mass of these $\Lambda_c^+$ candidates to the 
$\Lambda_c^+$ peak value using a
kinematic fitting program, and combine them with an anti-proton candidate 
and a number of pion candidates. We define the beam-constrained
mass as 
$M_B=\sqrt{E^2_{\rm beam}-(\Sigma_ip_i)^2}$, where $p_i$ is the 3-momentum vector for the
$i^{th}$  daughter of the $B$ candidate and $E_{\rm beam}$ is the beam energy. The resolution 
in $M_B$ is dominated by the spread in the CESR beam energy and is much better than
the resolution in the invariant mass of the combination.

For each combination, we calculated the energy difference 
$\Delta E=E_{\rm meas}-E_{\rm beam}$, 
where $E_{\rm meas}$ is the measured energy of the combination. A correctly reconstructed
$B$ meson has a $\Delta E$ distribution with a maximum at $0\ {\rm GeV}$. 
The $\Delta E $ resolution, $\sigma_{\Delta E}$, 
was calculated for each mode and 
detector configuration separately using the Monte Carlo simulation program, and combinations
were required to have $|\Delta E |<2 \sigma_{\Delta E }$. A further 
reduction in background is achieved by cutting on $\Theta_B$, the polar
angle of the $B$ in the laboratory frame with respect to the $e^+e^-$ axis. 
The distribution of ${\rm cos} \Theta_B$ is proportional to $ {\rm sin}^2\Theta_B$ for 
$e^+e^-\rightarrow\Upsilon(4S)\rightarrow B\overline{B}$, whereas background 
events are distributed 
nearly isotropically. We require $|{\rm cos} \Theta_B|<0.9$. If there are multiple candidates 
in an event with $M_B> 5.2\ {\rm GeV}/c^2$ for a given decay channel, 
the entry with the smallest
absolute value of $\Delta E$ is selected.

\begin{figure}[ht]
\begin{center}
\psfig{file=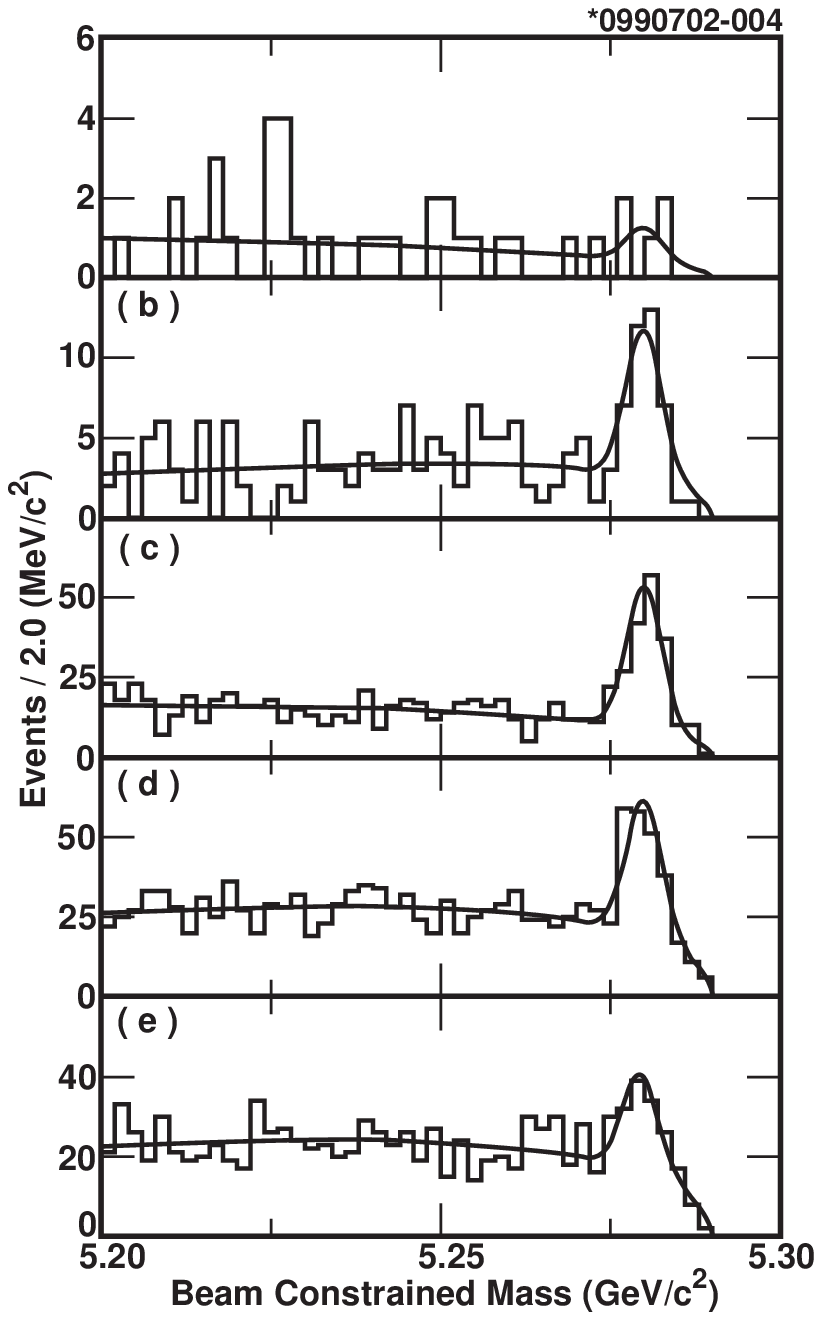,bbllx=2,bblly=-13,bburx=240,
bbury=385,width=4.0in}
\caption{Beam-constrained mass distributions for 
a) $\Lambda_c^+\overline{p}$,
b) $\Lambda_c^+\overline{p}\pi^-$,
c) $\Lambda_c^+\overline{p}\pi^-\pi^+$,
d) $\Lambda_c^+\overline{p}\pi^-\pi^+\pi^-$,
e) $\Lambda_c^+\overline{p}\pi^-\pi^0$.
}
\label{fig:mb}
\end{center}
\end{figure}

The $M_B$ distributions, after all selection criteria have been applied, 
are displayed in Fig.~\ref{fig:mb} for all modes investigated.
Strong signals are found in the modes $\Lambda_c^+\overline{p}\pi^-$ and 
$\Lambda_c^+\overline{p}\pi^-\pi^+$, confirming the previous observation of these modes.
Signals are also found in the new modes
$\Lambda_c^+\overline{p}\pi^-\pi^+\pi^-$, which has a
$13.0\sigma$ significance\footnote{We define our significance as the
probability, expressed in normal distribution sigma, of our
expected background to fluctuate to our signal's central
value. Poisson (Gaussian) statistics are used for expected
backgrounds with less (greater) than 30 events.}, and  $\Lambda_c^+\overline{p}\pi^-\pi^0$, which has a 
$8.2\sigma$
significance. There is no statistically significant signal in the two-body decay
$\overline{B^0}\rightarrow\Lambda_c^+\overline{p}$. Each $M_B$ distribution
is fit to a fixed width Gaussian signal function, and a background function  
of an exponential 
with phase-space threshold suppression. The
signal yields from these fits are shown in Table~\ref{table:yield}, 
where the uncertainties are 
statistical only. We have verified that similar distributions made with $\Lambda_c^+$
sidebands, $\Delta E $ sidebands, or continuum data show no peaking in the $B$ mass region.

Knowledge of the substructure of the multi-particle final states is very important.
From a purely practical point of view, the substructure changes the efficiency for
detecting a final state. This is particularly true when the intermediate 
particles are $\Lambda_{c1}$ and $\Sigma_c$ baryons which in turn decay strongly
with low $Q^2$ decays to $\Lambda_c^+$ baryons.
Furthermore, knowledge of the substructure gives 
information on the underlying mechanisms involved.

\begin{figure}[ht]
\begin{center}
\psfig{file=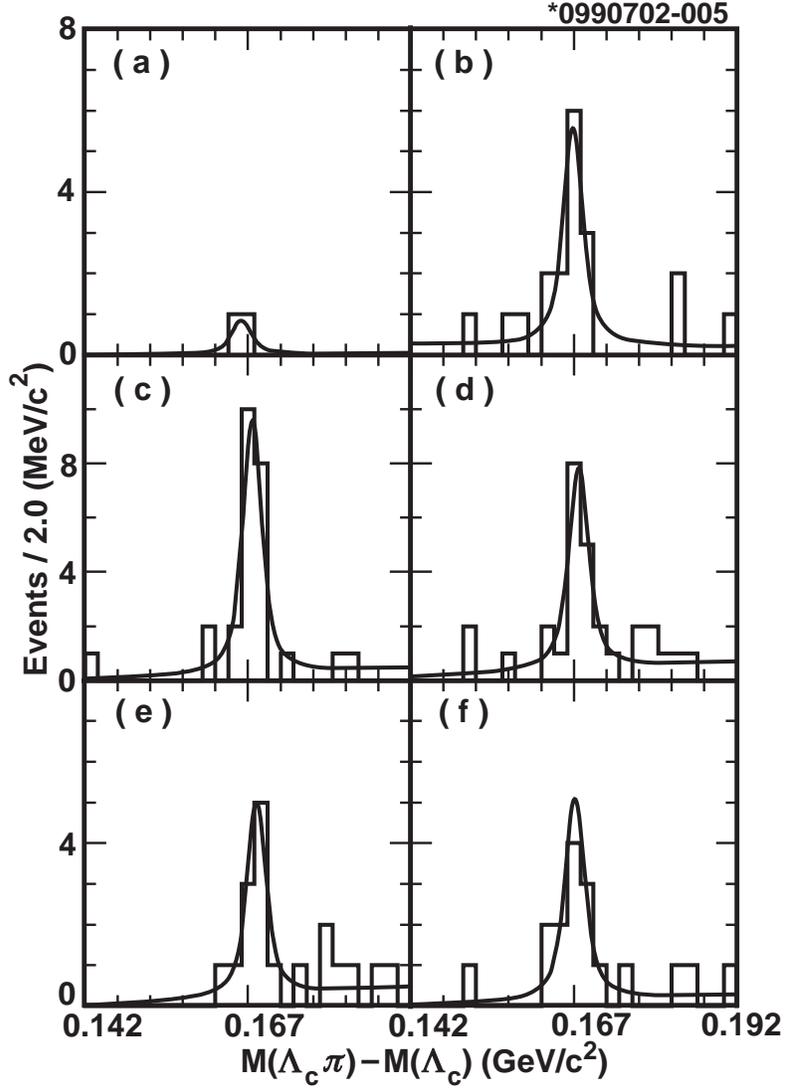,bbllx=2,bblly=-13,bburx=240,
bbury=385,width=4.0in}
\caption{
$M(\Lambda_c^+\pi)$-$M(\Lambda_c^+)$ mass differences 
for combinations within $2 \sigma$ of the 
B peak in the $M(B)$ distribution.
a) $M(\Lambda_c^+\pi^-)-M(\Lambda_c^+)$ within
$\Lambda_c^+\overline{p}\pi^-$,
b) $M(\Lambda_c^+\pi^-)-M(\Lambda_c^+)$ within
$\Lambda_c^+\overline{p}\pi^-\pi^+$,
c) $M(\Lambda_c^+\pi^+)-M(\Lambda_c^+)$ within
$\Lambda_c^+\overline{p}\pi^-\pi^+$,
d) $M(\Lambda_c^+\pi^-)-M(\Lambda_c^+)$ within
$\Lambda_c^+\overline{p}\pi^-\pi^+\pi^-$,
e) $M(\Lambda_c^+\pi^+)-M(\Lambda_c^+)$ within
$\Lambda_c^+\overline{p}\pi^-\pi^+\pi^-$,
f) $M(\Lambda_c^+\pi^-)-M(\Lambda_c^+)$ within
$\Lambda_c^+\overline{p}\pi^-\pi^0$.
}
\label{fig:sc}
\end{center}
\end{figure}

To search for $\Sigma_c^{++}\rightarrow\Lambda_c^+\pi^+$ and 
$\Sigma_c^0\rightarrow\Lambda_c^+\pi^-$,
we require that the combination 
have a beam constrained mass within $2\sigma_{M_B}$ 
of the $B$ mass peak to select
events in which a $B$ meson decays to a $\Lambda_c^+$. We then combine
this $\Lambda_c^+$ with a charged pion daughter of the $B$ decay and plot the
$M(\Lambda_c^+\pi)-M(\Lambda_c^+)$ mass difference (Fig.~\ref{fig:sc}). 
We fit these distributions
with a Breit-Wigner function of width defined by the CLEO measurements of the 
$\Sigma_c$ widths \cite{SIGWID}, convoluted with a Gaussian resolution function 
obtained from Monte Carlo studies, together with a polynomial background function.
We find good evidence for both $\Sigma_c^{++}$ and $\Sigma_c^0$
production in $\Lambda_c^+\overline{p}\pi^+\pi^-$ and
$\Lambda_c^+\overline{p}\pi^+\pi^-\pi^+$, and for $\Sigma_c^{0}$ in
$\Lambda_c^+\overline{p}\pi^0\pi^-$. All these signals have a
statistical significance greater than $5\sigma$.
Using analogous plots of those combinations
in the $M_B$ distributions outside the $B$ mass peak, we find negligible background
from true $\Sigma_c$ baryons that are not the daughters of the $B$ decay mode in question.
For the $\Sigma_c^0\overline{p}$ mode there are two events in the
signal region which would suggest a branching fraction of the order of
$0.25 \times 10^{-4}$. Our expected background in this mode, which is 0.12
events, has a 0.6\% chance of fluctuating to the 
observed events. We feel that this significance, is not sufficient to claim to have 
found a signal in this mode and we prefer to present a 90\% confidence level upper limit.
There are no events consistent with 
the production of $\Lambda_{c1}(2593)\to\Lambda_c^+\pi^+\pi^-$ or 
$\Lambda_{c1}(2625)\to\Lambda_c^+\pi^+\pi^-$ in these
decays which allows us to calculate 90\% confidence level upper limits on their production.

\begin{small}
\begin{center}
\begin{table}[ht]
\caption{ The yields for each mode. The events in the substructure are a 
subset of those in the main modes.}
\begin{tabular}{cccc}
Mode  &\hfil  Substructure \hfil & Total Yield & Substructure Yield \\
\hline
\hline
          $\Lambda_c^+\overline{p}$             & & $<8$     &  \\
\hline
          $\Lambda_c^+\overline{p}\pi^-$ \hfill & & $31\pm7$ & \\
& $ \Sigma_c^0\overline{p}$ & & $< 5.3$ \\
\hline
 $\Lambda_c^+\overline{p}\pi^-\pi^+$            & & $110\pm16$ & \\
&  $ \Sigma_c^0\overline{p}\pi^+$ \hfill        & & $14\pm4$ \\
&  $ \Sigma_c^{++}\overline{p}\pi^-$ \hfill     & & $23\pm5$ \\
&  $\Lambda_{c1}^+\overline{p}$ \hfill          & & $<2.3$  \\
\hline
 $\Lambda_c^+\overline{p}\pi^-\pi^+\pi^-$ & & $114\pm18$ & \\
&$ \Sigma_c^0\overline{p}\pi^+\pi^-$ \hfill    & & $19\pm5$ \\
&$ \Sigma_c^{++}\overline{p}\pi^-\pi^-$ \hfill & & $12\pm4$ \\
&$ \Lambda_{c1}^+\overline{p}\pi^-$ \hfill& & $<2.3$  \\
\hline
 $\Lambda_c^+\overline{p}\pi^-\pi^0$    & & $76\pm16$ & \\
&$ \Sigma_c^0\overline{p}\pi^0$ \hfill  & & $13\pm4$ \\

\end{tabular}
\label{table:yield}
\end{table}
\end{center}
\end{small}

\begin{table}[htb]
\caption{The efficiencies and branching fractions or 90\% CL upper
limits for each mode. 
Systematic uncertainties have been included in the upper
limits.  }
\begin{tabular}{cccc}
Mode     & Efficiency (\%)      & ${\cal B}$ ($10^{-4}$) & 
Previous Result ($10^{-4}$)\cite{CLEOO} \\
\hline
\hline
$\Lambda_c^+\overline{p}$      & 14.9  & $<0.9$                    & $<2.1$  \\
\hline
$\Lambda_c^+\overline{p}\pi^-$ & 15.8  & $2.4\pm0.6^{+0.19}_{-0.17}\pm0.6$ & $6\pm3$ \\
$\ \ \Sigma_c^0\overline{p}$   & 10.0  & $<0.8$&\\
\hline
$\Lambda_c^+\overline{p}\pi^-\pi^+$        & 12.7  &$16.7\pm1.9^{+1.9}_{-1.6}\pm4.3$ &$13\pm6$ \\
 $\ \ \Sigma_c^0\overline{p}\pi^+$ & 8.0& $2.2\pm0.6\pm0.4\pm0.5$ &  \\
 $\ \  \Sigma_c^{++}\overline{p}\pi^-$ & 7.8& $3.7\pm0.8\pm0.7\pm0.8$& \\
$\ \ \Lambda_{c1}^+\overline{p}$   & 3.2  &    $<1.1$                &\\
\hline
$\Lambda_c^+\overline{p}\pi^-\pi^+\pi^-$ & 9.3 & $22.5\pm2.5^{+2.4}_{-1.9}\pm5.8$   & $<15$ \\
 $\ \  \Sigma_c^0\overline{p}\pi^+\pi^-$ & 5.4 & $4.4\pm1.2\pm0.5\pm1.1$ & \\
 $\ \  \Sigma_c^{++}\overline{p}\pi^-\pi^-$ & 5.3 &
$2.8\pm0.9\pm0.5\pm0.7$ &  \\
$\ \ \Lambda_{c1}^+\overline{p}\pi^-$   & 1.9  &    $<1.9$                &\\
\hline
$\Lambda_c^+\overline{p}\pi^-\pi^0$ & 6.8 & $18.1\pm2.9^{+2.2}_{-1.6}\pm4.7$ & $<31$ \\
 $\ \  \Sigma_c^0\overline{p}\pi^0$ & 3.8 & $4.2\pm1.3\pm0.4\pm1.0$ &  \\
\end{tabular}
\label{table:effic}
\end{table}

Table~\ref{table:yield} 
shows the final results for the yields and branching fractions for all the 
modes. The efficiencies are calculated by our Monte Carlo simulation program. In this 
simulation, the 
$\Lambda_c^+$ decays were generated only into the three decay modes reconstructed,
using the measured branching ratios. To convert the quoted efficiencies to efficiencies 
which include the branching fractions of these modes,
they need to be multiplied by the absolute branching fraction of 
$\Lambda_c^+\rightarrow pK^-\pi^+$ of $5.0\pm1.3\%$\cite{PDG}.
The yield from the $\Sigma_c$ decay modes has been subtracted from the non-resonant
yields so that the resonant and non-resonant components can have different 
efficiency corrections applied. 

Table~\ref{table:yield} includes systematic uncertainties. Major contributors 
to these are uncertainties due to fitting
techniques, and uncertainties due to the efficiency calculation.
We take the fitting technique uncertainty as the maximum difference
obtained from different fitting methods. These techniques included using a scaled
$M_B$ distribution from $\Delta E $ sidebands for the background function, and 
fitting the $\Delta E $ distribution directly having first selected the $B$ mass
in the $M_B$ distribution. The uncertainties from this source are 5-17\%, 
dependent upon the mode.
The uncertainty in the efficiency calculation is 5-8\% due to uncertainties in the 
detection of the charged and neutral particles. In addition, there is a difference 
in efficiency due to possible substructure such as 
$\Lambda_c^*$, $\Sigma_c^*, \rho$ and $\Delta$
intermediate states. These all give a slightly reduced efficiency and thus give
an asymmetric systematic uncertainty. The systematic 
uncertainty due to the uncertainty in the $\Lambda_c^+\to pK^-\pi^+$ 
branching fraction is expressed as a third uncertainty.

Our limit on the branching fraction of the two body decay 
$\overline{B^0} \rightarrow \Lambda_c^+\overline{p}$ 
is $0.9 \times 10^{-4}$ at the 90\% confidence level. This 
is tighter than the previous CLEO limit of $2.0 \times 10^{-4}$. A recent theoretical
treatment by Cheng and Yang \cite{CY} using a bag model
predicts a branching fraction of this order, whereas older theoretical 
predictions\cite{OLDER} predicted larger numbers by a factor of at least
4 over our experimental limit. Our measurement of
the three-body 
branching fraction $B^- \rightarrow \Lambda_c^+\overline{p}\pi^-$ and
$\overline{B^0} \rightarrow \Lambda_c^+\overline{p}\pi^-\pi^+$ 
are both consistent with, and much more accurate than, 
the previously measured modes.
The three three-body decays $\overline{B^0} \rightarrow \Sigma_c^{++}\overline{p}\pi^-$,
$\overline{B^0} \rightarrow \Sigma_c^{0}\overline{p}\pi^+$ and $\overline{B^-} \rightarrow
\Sigma_c^{0}\overline{p}\pi^0$ have essentially identical
phase-space, but only the $\Sigma_c^{++}$ decay can proceed via both
external and internal $W$ decay diagrams, whereas 
the $\Sigma_c^{0}$ decays can only proceed via an internal $W$. We
find the rate of all three decays to be of the same order.
This implies that the external $W$ decay diagram does not dominate over the 
internal $W$ decay 
diagram, although naively we would expect the latter to be color-suppressed.

In conclusion, we have measured branching fractions of $B$ mesons into the decay modes
$\Lambda_c^+\overline{p}\pi^-$,
$\Lambda_c^+\overline{p}\pi^-\pi^+$,
$\Lambda_c^+\overline{p}\pi^-\pi^+\pi^-$, and 
$\Lambda_c^+\overline{p}\pi^-\pi^0$. 
The first two of these confirm, with greater accuracy, 
the previous measurements. 
The latter two are the first observations of these decay modes.
We find a limit on the two body decay $\overline{B^0} \rightarrow \Lambda_c^+\overline{p}$, which 
discriminates between theoretical models. We make the first measurements of 
exclusive states that include $\Sigma_c^{++}$ or $\Sigma_c^0$ baryons. Our 
measurements indicate that external $W$ diagram decays do not dominate
over the 
competing internal $W$ diagram decays in Cabibbo-favored baryonic B decays.

We gratefully acknowledge the effort of the CESR staff in providing us with
excellent luminosity and running conditions.
M. Selen thanks the PFF program of the NSF and the Research Corporation, 
and A.H. Mahmood thanks the Texas Advanced Research Program.
This work was supported by the National Science Foundation, and the
U.S. Department of Energy.

\end{document}